\documentclass[aip,amsmath,amssymb,showpacs,nofootinbib,reprint]{revtex4-1}
\usepackage{amsmath}
\usepackage{graphicx}
\usepackage{subfigure}
\usepackage{lipsum}
\usepackage{bm}
\newcommand{\kpar}{k_{\parallel}}

\newcommand{\om}{\omega}
\newcommand{\lk}{\left (}
\newcommand{\rk}{\right )}
\newcommand{\lK}{\left [}
\newcommand{\rK}{\right ]}

\newcommand{\ekx}{e^{-k_{\parallel}\vert x-x^{\prime}\vert}}
	
\newcommand{\xp}{x^{\prime}}

\newcommand{\be}{\begin{equation}}
\newcommand{\ee}{\end{equation}}	
\newcommand{\bea}{\begin{eqnarray}}
\newcommand{\eea}{\end{eqnarray}}
\newcommand{\idd}{\int_{\frac{d}{2}}^{\frac{d}{2}}}

\newcommand{\bnabla}{\boldsymbol{\nabla}}
\newcommand{\ul}{\underline{}}				
\begin{document}

\title{Long-wave magnons in a ferromagnetic film}
\author{Gang Li}
\email{dgzy03@tamu.edu}
\affiliation{ Department of Physics, Texas A\&M University, College Station, Texas 77843-4242, USA}
\author{Chen Sun}
\affiliation{ Department of Physics, Texas A\&M University, College Station, Texas 77843-4242, USA}
\author{Thomas Nattermann}
\affiliation{ Institute f\"ur Theoretische Physik, Universit\"at zu K\"oln, Z\"ulpicher Str. 77a, D-50937 K\"oln,
			 Germany}
\author{Valery L. Pokrovsky}
\affiliation{ Department of Physics, Texas A\&M University, College Station, Texas 77843-4242, USA}
\affiliation{Landau Institute for Theoretical Physics, Chernogolovka, Moscow District, 142432, Russia}
\begin{abstract}
	An asymptotically exact theory of spectrum and transverse distribution of magnetization in long-wave magnons is presented. 
	It is based on exact analytical solution of linearized Landau-Lifshitz equation in a film. The quantization of the transverse 
	wave vector and role of evanescent waves at different values of parameters and wave vectors is studied.
\end{abstract}
\maketitle

\noindent\textbf{Introduction.}
In this article we present asymptotically exact theory of the spectrum
and transverse distribution of magnetization in long-wave length  magnons
propagating in a ferromagnetic film. The theory is based on
exact solution of linearized Landau-Lifshitz equation (LLE). To avoid
complications we assume the film to be isotropic in the film  plane, and external magnetic field $H$ and the spontaneous magnetization $M$ to be oriented
in plane. Their direction is accepted for $z-$axis, whereas the $x-$axis
is directed perpendicular to the film that occupies the volume between
parallel planes $x=\pm\frac{d}{2}$. The Hamiltonian $\mathcal{H}$
includes exchange, Zeeman and dipolar interaction of magnetization:
\begin{eqnarray}\label{eq:Hamiltonian}
\mathcal{H}&=&\int_V d^3r\Bigg[\frac{D}{2}\left(\nabla_\alpha \mathbf  {M}\right)^{2}-{\mathbf H} \cdot {\mathbf M}\nonumber\\
&+&\frac{1}{2}\left({\mathbf M}\cdot \boldsymbol{\nabla}\right)\int_V d^3r'\left(\mathbf  {M'}\cdot\boldsymbol{\nabla}'\right)\frac{1}{\left|\mathbf {r-r'}\right|}\Bigg]
\end{eqnarray}
Here $\mathbf {M\left(r\right)}$ is the local magnetization vector, 
$D$ is the exchange energy divided by  $M^{2}a$, where $a$
is the lattice constant. $V$ stands for volume and prime denotes dependence on the coordinate $\mathbf {r'}$. Note that
the coefficient $D$ has dimensionality of square of length. The value $\ell=\sqrt{D}$ is called dipolar length. This is the scale of
distance at which dipolar and exchange interactions become of the
same order of magnitude. Typically it is about 10-30nm.
Theory of magnons in ferromagnetic films has important applications
to real magnets and rather long history. In this brief article we
can cite only several most important articles and give our apologies
to many contributors to the subject. The first exact result was obtained
by Damon and Eshbach \cite{Damon 1961} for purely dipolar interaction.
Kalinikos \cite{Kalinikos 1980} and Kalinikos and Slavin \cite{Kalinikos 1986}
incorporated exchange interaction together with dipolar one and obtained
an integral-differential equation for magnetization in a spin wave.
They solved it employing a plausible, but uncontrollable approximation.
Rezende \cite{Rezende 2009} assumed that magnetization does not depend
of transverse coordinate $x$ and exactly diagonalized the resulting
Hamiltonian. Though such assumption is qualitatively justified for the
transverse mode with the lowest energy, it is obviously invalid for
higher transverse modes. In a recent work Sonin \cite{Sonin 2017}
has found the magnon spectrum and shape of transverse mode at zero
$y-$component of the magnon wave vector $k_{y}$. His solution allows
an explicit analytical expression in the limit $d\gg \ell$ and 
$1/d\ll k_z\ll 1/\ell$. 
Our work is an extension of Sonin's method to the case of general $k_{y}$.

We show that formally the spectrum of magnons in a film has the same
analytical form as in the bulk, but quantization of the transverse
wave vector and transverse distribution of magnetization depend on
thickness of the film $d$ and other variables in a highly non-trivial
way. In contrast to standard semiclassical approximation that
becomes valid when the number of transverse mode $n$ is a large number,
in the ferromagnetic film if $d\gg \ell$ there exist two different 
asymptotics in the ranges $1\ll n\ll d/\ell$ and $n\gg d/\ell$. In
both cases for each $n$ the distribution of magnetization across the 
film consists of one oscillating mode and two evanescent modes. All they have the same frequency. We obtain analytical solution at $d\gg\ell$ for any $n$, not only
large ones. Due to symmetry of the problem magnon spectrum is divided in  two
infinite series. In simplest situation they correspond to even and odd transvese distribution of magnetization, $n$-th mode oscillates $n$ times between boundaries. 
Evanescent waves can be neglected in the exchange-dominated range of wave vectors
$\kpar\equiv\sqrt{k_y^2+k_z^2}\gg 1/\ell$. Otherwise evanescent waves must be taken in account. Specifically, they play important role
in the case of thin films $d \leq \ell$. 
Theory for small linear sizes has a special interest for applications to devices employing magnons instead of electrons as carriers of information.\cite{Sun 2014,Duine 2015}
Our method can be extended to include
anisotropy (spin-orbit interaction), tilted external magnetic field
and other shapes of the sample.

\noindent\textbf{Equations of motion and magnon solutions.}
A weak excitation of the equilibrium state is described by the transverse
components of magnetization ${\mathbf M}\equiv \lk M_{x},M_y\rk$.
They obey the linearized LLE:
\begin{eqnarray}
\dot{{\mathbf M}}&=&\gamma\lK \left(H-MD\Delta\right){\mathbf M}+M {\mathbf h}\rK\times\hat z, 
\label{eq:linearized}
\end{eqnarray}
$\hat z$ is the unit vector in $z$-direction and  $\Delta\equiv\bnabla^2$; ${\mathbf h}=\bnabla_\perp \phi$ denotes the dipolar field induced by magnetization inside and outside the film, with  $\bnabla_\perp\equiv\left(\begin{matrix}\partial_x,\partial_y\end{matrix}\right)^\top$ and
\begin{equation}\label{eq:h-phi}
\phi\left(\mathbf{r}\right)=-\bnabla_\perp \cdot \int d^3r'{\mathbf M}'\left|\mathbf {r-r}'\right|^{-1}.
\end{equation}
The number of parameters of the present problem can be reduced by the scale transformations
\begin{equation}\label{eq:phi-eta}
t \to \omega_H^{-1}t,\qquad \mathbf  r\to \sqrt{\frac{\chi}{4\pi}}\,\ell\,\mathbf  r,\qquad\mathbf M\to M\mathbf M.
\end{equation}
Here $\omega_H\equiv\gamma H$ denotes the Larmor frequency and $\chi\equiv 4\pi M/H$ the static magnetic susceptibility (we absorb a factor $4\pi$ in its definition to simplify final expressions). 
In rescaled units  equation of motion simplifies to 
\begin{equation}
\label{LLE2}
\dot{{\mathbf M}}=\lK\left(1-\Delta\right) {\mathbf M}+ \frac{\chi}{4\pi}  {\mathbf h}\rK\times \hat z . 
\end{equation}
The equations for ${\mathbf h}$ and $\phi$ remain unchanged. The remaining parameters of the problem are susceptibility $\chi$ and half of the sample width $d/2$ in new units

Applying laplacian $\Delta$ 
to eq. (\ref{LLE2}) and using magnetostatic equation
\begin{equation}\label{phi2}
\Delta\phi=4\pi\bnabla_\perp\cdot{\mathbf M},
\end{equation}
one gets the desired equation for ${\mathbf M}$:
\begin{equation}
\Delta\dot{{\mathbf M}}=\lK\lk\! 1-\Delta\rk\!\Delta{\mathbf M}+\chi \bnabla_\perp\!\lk\bnabla_\perp\!\cdot\!{\mathbf M}\rk  \rK\!\times\!\hat z.\label{LLE3} 
\end{equation}
It must be solved with standard boundary conditions (BC) for Maxwell equations that requires continuity of tangential components of magnetic field $\mathbf h$ and normal component of magnetic induction $\mathbf{b}=\mathbf{h}+4\pi\mathbf{M}$ at two surfaces of the film. Another set of BC originates from variation of magnetization (spins) on surfaces if they are free. It leads to equations:
\begin{equation}\label{boundary-exch}
\partial_x \mathbf M\big|_{x=\pm d/2}=0.
\end{equation}
We call them exchange boundary conditions (EBC).
 
In a propagating wave with in-plane wave vector $\mathbf {k_{\Vert}}=k_{y}\hat{y}+k_{z}\hat{z}$, 
the oscillating components of magnetization can be written as 
\begin{equation}\label{solution 0}
{\mathbf M} = \left(\begin{matrix} m_x(x)\cos\lk\mathbf {k_{\Vert}\cdot r}-\omega t\rk\\m_y(x)\sin\lk\mathbf {k_{\Vert}\cdot r}-\omega t\rk\end{matrix}\right).
\end{equation}
The Ansatz (\ref{solution 0}) turns  eq. (\ref{LLE3}) into 
a system of ordinary linear homogeneous differential equations with constant coefficients for the  vector field ${\mathbf m}\equiv \lk m_x,m_y\rk^\top$ which  describes the transverse distribution of magnetization.      
General solution of such a system is a superposition of basic exponential solutions $ {\mathbf m}(x)={\mathbf m}_0 e^{ik_xx}$. After division by $k^2=k_\parallel^2+k_x^2$  equation for  ${\mathbf m}_0$ reads  
\begin{equation}\label{eq:m}
\lk  \underline\Omega -iB\underline\sigma_3\rk\mathbf m_0=0,\qquad  \underline\Omega=\left(\begin{matrix}\omega&-A_y\\-A_x&\omega\end{matrix}\right).
\end{equation} 
Here  $A_{\alpha}=1+k^2+{\chi}\hat k_{\alpha}^2$,   and $B={\chi}\hat k_x\hat k_y$.  $\hat k_\alpha=k_\alpha/k$ denotes the cosine of direction and $\underline\sigma_3$ is the  
Pauli matrix. The solvability condition of eq. (\ref{eq:m}),  $\omega^2+B^2-A_xA_y=0$, delivers   the magnon dispersion relation: 
\begin{equation}
\omega^2=\lk 1+k^{2}\rk  \lk 1+\chi+k^{2}-\chi  \hat k_{z}^{2}\rk .\label{dispersion}
\end{equation}
It  does not depend on the sample thickness  and has therefore
the same form as in the bulk. Boundary conditions will however restrict possible $\mathbf k$-vectors, as it will be shown below. The dispersion relation (\ref{dispersion}) can be treated as a cubic equation for $k^2$, assuming that $\omega$ and $k_z$ are given. Its  three solutions 
can be written as  $k_i^2=k_{x,i}^2+k_\parallel^2$. Thus $k_{x,i}$ is a function of $\omega$ and $\mathbf k_\parallel$. Close investigation shows that all 3 roots of cubic equation for $k^2$ are real,
one of them $k_{1}^2$ is positive, two others $k_{2}^2$ and $k_{3}^2$ are negative in the entire physically available range of parameters. Positive root $k_1^2$ corresponds to oscillating transverse mode, two negative roots correspond to evanescent waves. 

Equations (\ref{eq:m}) and boundary conditions are invariant under operation $x\rightarrow -x, k_y\rightarrow -k_y$. It means that 
all eigenvaliues $\omega$ are at least double degenerate and the eigenfunctions with the same $\omega$ and opposite signs of $k_y$
are connected with a simple relation:
\begin{equation}\label{symmetry}
m_{x,y}(x;k_y)=m_{x,y}(-x;-k_y).
\end{equation}
The value $k_z$ enters in equations only as $k_z^2$. Therefore, the solution does not change at transformation $k_z\rightarrow -k_z$. 
These properties can be obtained from invariance of the Hamiltonian with respect to two discrete transformations: reflection in the central plane
of the film combined with time reversal and reflection in the $(x,z)$-plane combined with time reversal. Time reversal
in addition to reflection is necessary to keep pseudo-vector of spontaneous magnetization invariant.

The transverse distribution of magnetization $\mathbf{m}(x)$ must be a real vector field. Therefore for any mode it can be written as follows:
\begin{equation}\label{eq:real}
\mathbf{m}(x) = \mathbf{a}\cos k_x x + \mathbf{b}\sin k_x x,
\end{equation}
where $\mathbf a$ and $\mathbf b$ are real constant vectors. According to (\ref{eq:m}), the coefficients $\mathbf a, \mathbf b$ obey the relation  $\underline\Omega\cdot \mathbf a - B\underline\sigma_3\mathbf b=0$ 
which implies the  amplitude relations  
\begin{equation}\label{amplitude relations}
\mathbf b = \underline \Lambda (\mathbf k,\omega)\cdot \mathbf a,\qquad \ul\Lambda=\ul\sigma_3 \ul\Omega/B.
\end{equation}
Symmetry discussed above retains invariant coefficents $\mathbf{a}_i$ and changes sign of coefficients $\mathbf{b}_i$ ($i=1,2,3$). 

\noindent\textbf{Boundary conditions and consistency requirement}.  
The exchange BC (\ref{boundary-exch}) include 4 equations, two on each surface. 
They cannot be satisfied with a single-mode solution (\ref{eq:real}) associated with one of three posssible values of $k_x^2$. Indeed, according to eq. (\ref{amplitude relations}) such a solution depends only on two independent parameters, for example $a_x, a_y$. Only a proper superposition of three solutions can satisfy exchange and electromagnetic BC simultaneously. Such a general solution of the equation (\ref{eq:m}) represents the vector $\mathbf{m}(x)$ as a superposition: 
\begin{equation}\label{superposition}
\mathbf{m}(x)=\sum_{i=1}^{3}\left(\mathbf{a}_i\cos k_{ix}x+\mathbf{b}_i\sin k_{ix}x\right),
\end{equation}
where $k_{ix}$ denotes the $x$-component of the wave vector corresponding to $i-$th solution of cubic equation (it is purely imaginary for evanescent waves) and $\mathbf{a}_i,\mathbf{b}_i$ are the vector amplitudes of $i-$th mode. 
Using eq. (\ref{superposition}), the EBC 
equations can be rewritten in terms of 12 independent coordinates of vectors $\mathbf{a}_i,\mathbf{b}_i;\,\,i=1,2,3$:
\begin{equation}\label{eq:boundary-ex}
\sum_{i=1}^{3}a_{ix}k_{xi}\sin\alpha_i=0;\,\,\,
\sum_{i=1}^{3}b_{ix}k_{xi}\cos\alpha_i=0;\,\,\,\alpha_i= \frac{k_{ix}d}{2}.
\end{equation}

The magnetostatic BC are satisfied automatically for any distribution of magnetization if magnetic potential obeys the integral relation (\ref{eq:h-phi}). In particular, it will be satisfied for magnetization represented by superposition (\ref{superposition}). We have proved that any solution of equation of motions must be such a superposition. However, the inverse statement that any such a superposition is solution of equations of motions (\ref{LLE2}) is wrong. It happens because equations of motion contain not only differential, but also integral terms. The choice of valid solutions is realized by \textit{condition of consistency}. It requires
magnetic potential $\phi$ to be a superposition of exponents $e^{\pm ik_ix}$ where
$k_i^2$ are solutions of the cubic equation discussed earlier. We will see that integrals in $\phi (x)$ eq. (\ref{eq:h-phi}) generate extra exponents of the type $e^{\pm\kpar x}$ that are not allowed by cubic equation. Consistency requires coefficients at them to be zero. Below we display an explicit form of these consitency equations (CE).

Integration over longitudinal coordinates $y,z$ in eq. (\ref{eq:h-phi}) can be performed explicitly with the result:
\begin{equation}\label{phi-eta}
\phi = -4\pi\left(d_x\eta_x+k_y\eta_y\right);\,\,\,\eta_{j}=\frac{1}{2\kpar}\idd
\ekx m_j(\xp)d\xp.
\end{equation}
The basic integrals that enter $\eta_j(x)$, $j=x,y$ are:
\begin{equation}\label{eq:integrals}
\begin{aligned}
&\idd d\xp\frac{\ekx}{2\kpar}\left(\begin{matrix}\cos k_{ix}\xp\\\sin k_{ix}\xp\end{matrix}\right)           \\
=&\frac{1}{k_i^2}\left(\begin{matrix}\cos k_{ix} x\\\sin k_{ix} x\end{matrix}\right) -
\frac{e^{-\frac{\kpar d}{2}}}{\kpar k_i^2}\left(\begin{matrix}f_{ic}\cosh\kpar x\\f_{is}\sinh\kpar x\end{matrix}\right).
\end{aligned}
\end{equation}
Here $k_i^2=\kpar^2+k_{ix}^2$ and we denotes $f_{ic}=\kpar\cos\alpha_i-k_{ix}\sin\alpha_i$,
$f_{is}=\kpar\sin\alpha_i+k_{ix}\cos\alpha_i$ with $\alpha_i=k_{ix} d/2$. Eq. (\ref{eq:integrals}) visibly demonstrates the appearance in magnetic potential of exponents $\exp(\pm\kpar x)$ forbidden by secular cubic equations for $k^2$ since it corresponds to $k^2=0$. It vanishes in $\phi$ only due to superposition. The consistency equations require coefficients at $\cosh\kpar x$ and $\sinh\kpar x$ to be zero. The corresponding equations can be written as follows:
\begin{align}\label{eq:consistency-gen}
\sum_{i=1}^3& \frac{1}{k_i^2}\left(\kpar a_{ix}f_{ic}+k_yb_{iy}f_{is}\right)=0\nonumber\\
\sum_{i=1}^3& \frac{1}{k_i^2}\left(\kpar b_{ix}f_{is}+k_ya_{iy}f_{ic}\right)=0.
\end{align}
In order to turn CE together with exchange boudary conditions (\ref{eq:boundary-ex}) into closed system of 6 equations for 6 independent amplitudes $a_{ix},b_{ix}$ it is possible to use relations between $a_{iy},b_{iy}$ and $a_{ix},b_{ix}$ following from equations of motion in form (\ref{eq:m}):
\begin{equation}\label{eq:aby-abx}
a_{iy}=\frac{\om}{A_{iy}}a_{ix}-\frac{B_i}{A_{iy}}b_{ix};\,\,\,b_{iy}=\frac{B_i}{A_{iy}}a_{ix}+\frac{\om}{A_{iy}}b_{ix},
\end{equation}
where $A_{iy}=1+k_i^2+\frac{\chi k_y^2}{k_i^2}$ and $B_i=\frac{\chi k_{ix}k_y}{k_i^2}$. Besides of that it is necessary to eliminate values $k_i^2$ and $k_{ix}$ with $i=2,3$. As it follows from cubic equation for $z=k^2$, if the first (positive) root $z_1=k_1^2$ is fixed, two others can be found from equation:
\[k_{2,3}^2=-1-\frac{\chi}{2}-\frac{k_1^2}{2}\pm\sqrt{\left(1+\frac{\chi}{2}+\frac{k_1^2}{2}\right)^2-\frac{\chi k_z^2}{k_1^2}}.
\]
In this way all $k_i^2$ and $k_{ix}$ with $i=2,3$ are determined through the single
positive wave vector $k_{1x}$.  

The system of 4 EBC (\ref{eq:boundary-ex}) and 2 EC equations considered as 6 linear homogeneous equations for $a_{ix}, b_{ix}$ has solutions only if its determinant is equal to zero. This requirement determines discrete set of $k_{1x}$, i.e. transverse quantization of wave vector. This equation is exact in the framework of considered model. In Fig. 1 we show results of numerical calculations of quantized spectra from requirements of zero determinant for $d=100$, and $\chi=2$ for direction of propagation perpendicular  and parallel to magnetization and spectra of the first transverse modes for a few different directions of propagation.

To give a more visible idea of mathematical procedure leading to these results, we consider 6 EBC-CE equations in some detail. Only $\cos k_{1x} d/2$ and $\sin k_{1x} d/2$ are oscillating functions of their arguments. The functions $f_{1c},f_{1s}$ are linear functions of $\cos\alpha_1,\,\, \sin\alpha_1$ and therefore also oscillate. Other functions containing $\cos\alpha_{2,3},\,\,\sin\alpha_{2,3}$ are hyperbolic functions and change monotonically with $k_{1x}$. In the 6$\times$6 matrix $C$ of the EBC-CE equations the first two columns are linear combinations of $\sin\alpha_1,\,\,\, \cos\alpha_1$, the rest are monotonical functions. Therefore, the determinant has a form $\det C= {\cal K}\sin^2\alpha_1+2{\cal L}\sin\alpha_1\cos\alpha_1+{\cal M}\cos^2\alpha_1$, where $\cal{K,L,M}$ are monotonical functions of $k_{1x}$ or $\alpha_1$. Then equation $\det C=0$ can be rewritten as ${\cal K}\tan^2\alpha_1+2{\cal L}\tan\alpha_1+{\cal M}=0$ with formal solution:

\[ \tan\alpha_1=\frac{-{\cal L}\pm\sqrt{{\cal L}^2-\cal{KM}}}{2{\cal K}}.
\]

\
This is an implicit equation for $k_{1x}$. It shows that the consequent quantized values $k_1x$ are located between points $n\pi/d$ and that quantized values form two series corresponding to two signs in front of square root in previous equation. Thus, the quantized values of $k_{1x}$ can be enumerated by an index $\nu$ taking two values $+$ and $-$ and by an integer number $n$ taking values from 0 to $\infty$. We will denote these quantized values as $k_{x\nu n}$. For large $n$ the main part of $k_{x\nu n}$ is $2\pi n/d$. An approximate formula for the quantized values reads:
\begin{equation}\label{eq:quant-large-n}
k_{x\nu n}=\frac{2\pi n}{d}+\frac{2}{d}\arctan\frac{-{\cal L}\pm\sqrt{{\cal L}^2-\cal{KM}}}{2{\cal K}}.
\end{equation} 
In the argument of $\arctan$ $k_{1x}$ must be replaced by $2\pi n/d$. 
\onecolumngrid
\begin{center}
\begin{figure}
\subfigure[]
	{\includegraphics[width=5.6cm]{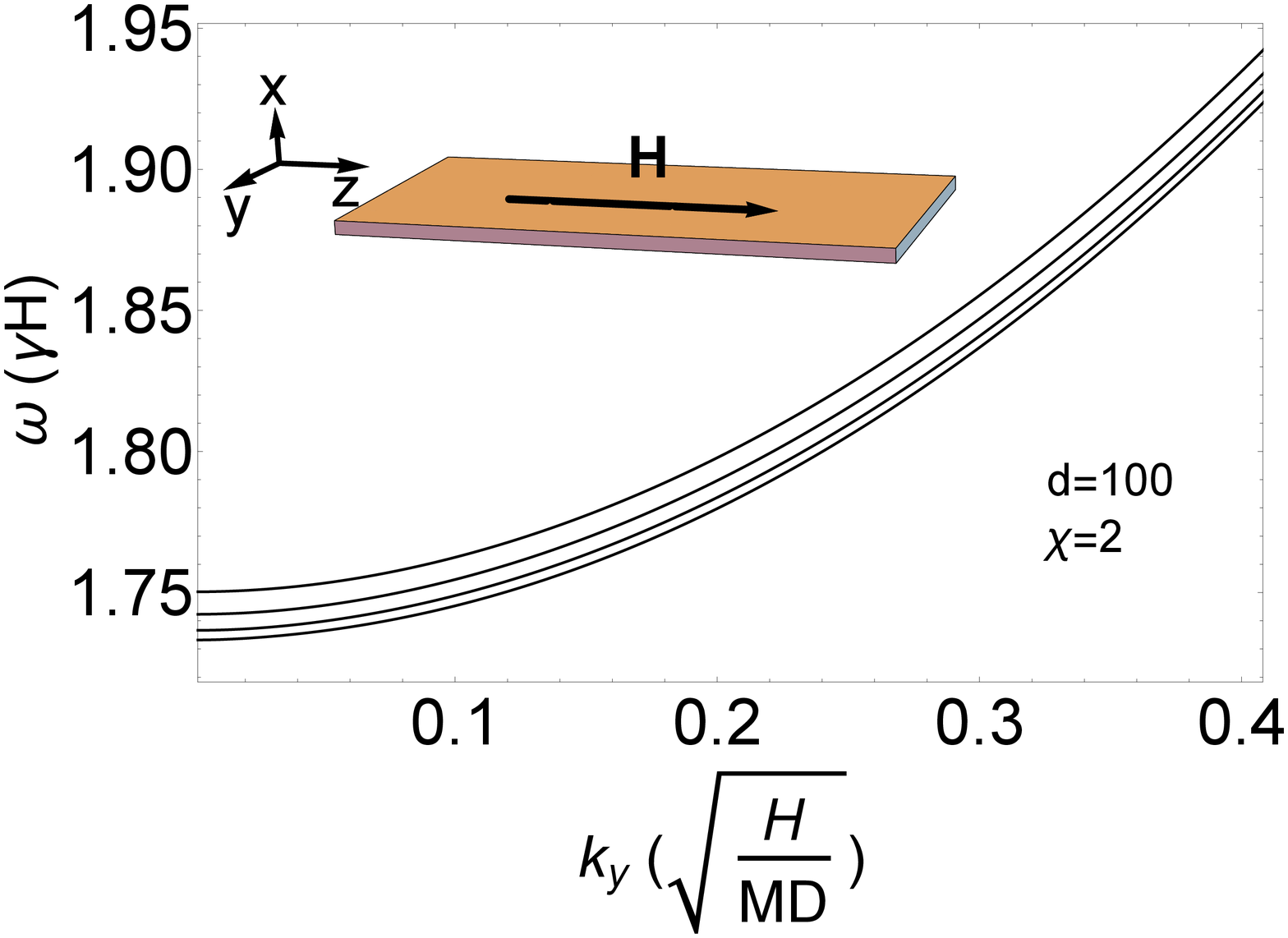}}	
\subfigure[]
	{\includegraphics[width=5.55cm]{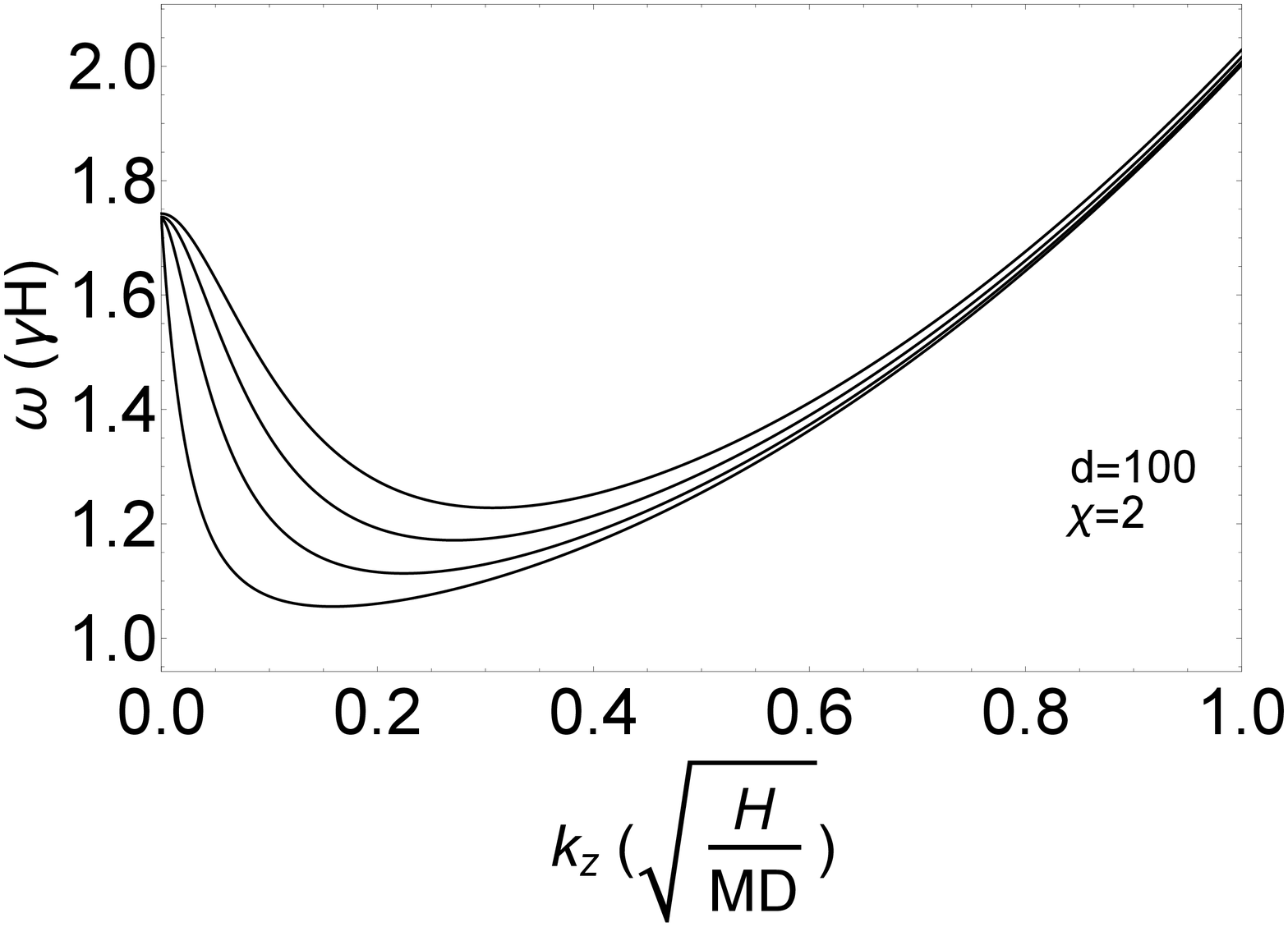}}
\subfigure[]
	{\includegraphics[width=5.42cm]{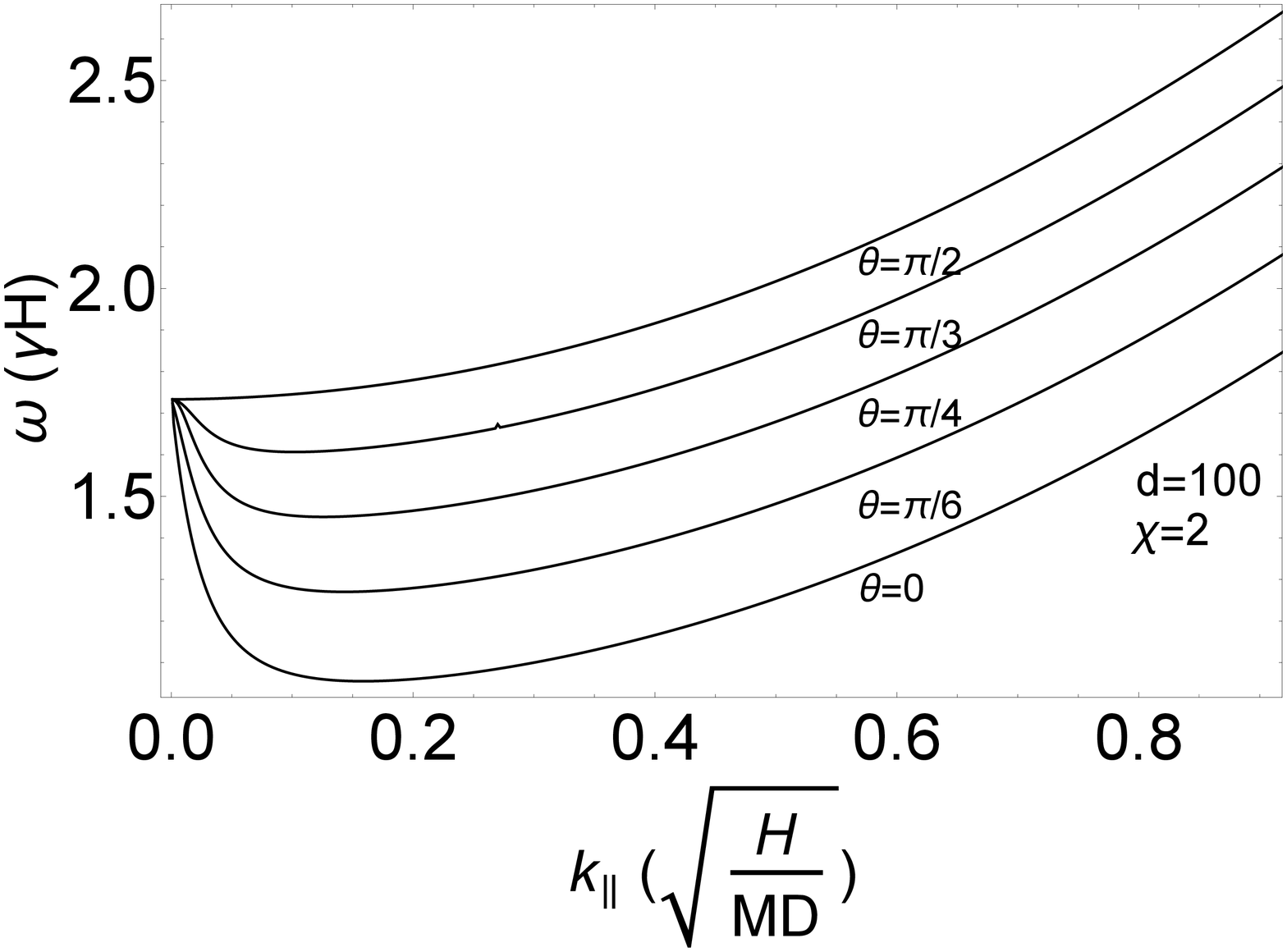}}
	\caption{Results of numerical calculations for the case $d=100$ and $\chi=2$. (a) The first four quantized spectra  for direction of propagation  perpendicular to magnetization.
	(b) The first four quantized spectra  for direction of propagation  parallel to magnetization.
	(c) Spectra of the first transverse modes for $\theta=0,\frac{\pi}{6},\frac{\pi}{4},\frac{\pi}{3} ,\frac{\pi}{2}$.}
\end{figure}
\end{center}
\twocolumngrid

\noindent\textbf{General properties of magnon spectra in thick films.}
In conclusion we describe general properties of spectra and structure of transverse modes in thick films $d\gg 1$.

At fixed direction of propagation given by $\theta\equiv\arccos(k_z/\kpar)=const$, the frequency $\om_{\nu n}$ of a mode $\nu n$ with $n\ll d/2\pi$ as function of $\kpar$ has a minimum at non-zero value 
\[k_{\parallel 0}\approx\left(\frac{\chi\cos^2\theta}{2+\chi\sin^2\theta}\right)^{1/4}k_{x\nu n}^{1/2}.\]
 According to this equation, $k_{\parallel 0}\gg k_{x\nu n}$, but it is much less than 1 that makes this explicit result available. The limitation to $n$ ensures that $k_{x\nu n}\ll 1$ and as a consequence $k_{\parallel 0}\ll 1$. The frequency in minimum is $\om_{min}=1+O(k_{\parallel 0}^2)$. The position of minimum eventually shifts to larger $\kpar$ with the growth of $n$ or equivalently $k_{x\nu n}$ and at $k_{x\nu n}=\frac{1}{2}\left(\sqrt{( 2+\chi\sin^2\theta)^2+8\chi\cos^2\theta}-2-\chi\sin^2\theta\right)^{1/2}$ goes to $+\infty$.
At fixed $n$ and $\theta$ decreasing, $k_{\parallel 0}$ decreases. At $\theta=0$ minimum and maximum coalesce. The point $k_y=0$ is the only
minimum of frequency in the spectrum of any magnon mode propagating perpendicularly to permanent magnetization ($\cos\theta=0$).

Maximum of frequency for any fixed $\theta$ except of $\theta=\pi/2$ is located at
$\kpar=0$. The value of frequency in maximum is equal to $\om_{max}=1+\chi$, frequency of ferromagnetic resonance. For a mode with $n$ not large such a value of frequency is reached right of minimum at $\kpar=\lk\sqrt{1+\chi+\chi^2\sin^2\theta/2}-1-\chi^2\sin^2\theta/2\rk^{1/2}$. This result shows a rather strong asymmetry of the spectral curve with respect to its minimum generated by dipolar forces. This asymmetry is very important for different applications. In particular, the presence of two minima is definitive in structure of magnon Bose-condensation and possible superfluidity .\cite{Chen 2017}

The transverse distribution of magnetization in each mode is described by superposition of one oscillating and two evanescent modes. Oscillating mode is a sum
of the type $a\cos k_x x + b\sin k_x x$. At any direction of propagation except of
$\theta=0$ both $a$ snd $b$ are not zero. The mode becomes even or odd only at $\theta=0$. However, the parity tends to be conserved at large $\kpar\gg 1$ when the exchange interaction dominates. In this case the dipolar interaction can be neglected, evanescent waves responsible for the balance of magnetic field and induction on the surface also are small. The EBC in this case are satisfied by even or odd transverse distribution of magnetization.

 \noindent\textbf{Acknowledgements.}We thank E.A. Sonin for sending us his manuscript prior publication and discussions. This work was supported by the University of Cologne Center of Excellence QM2.

\end{document}